\documentclass[fleqn,usenatbib]{mnras}

\usepackage{newtxtext,newtxmath}

\usepackage[T1]{fontenc}
\usepackage{ae,aecompl}

\usepackage{graphicx}	
\usepackage{amsmath}	
\usepackage{amssymb}	

\usepackage{color}

\newcommand{\Msun}{M$_\odot$}
\newcommand{\Mstar}{M\ensuremath{_\star}}
\newcommand{\lamRe}{\ensuremath{\lambda_\mathrm{Re}}}
\newcommand{\lamReeps}{\lamRe\,$-\,\varepsilon$}

\title[Evolution of \lamRe\ and its connection with mergers and gas accretion]{Kinematic analysis of EAGLE simulations: Evolution of \lamRe\ and its connection with mergers and gas accretion}

\author[D. Walo-Mart\'in et al.]{
D. Walo-Mart\'in$^{1,2}$\thanks{E-mail: dwalo@iac.es},
J. Falc\'on-Barroso$^{1,2}$,
C. Dalla Vecchia$^{1,2}$,
I. P\'erez$^{3,4}$, 
A. Negri$^{1,2}$
\\
$^{1}$Instituto de Astrof\'isica de Canarias, Calle V\'ia L\'actea s/n, E-38205 La Laguna, Tenerife, Spain\\
$^{2}$Departamento de Astrof\'isica, Universidad de La Laguna, Av. del Astrof\'isico Francisco S\'anchez s/n, E-38206, La Laguna, Tenerife, Spain\\
$^{3}$Departamento de F\'isica Te\'orica y del Cosmos, Universidad de Granada, Facultad de Ciencias (Edificio Mecenas), E-18071, Granada, Spain\\
$^{4}$Instituto Carlos I de F\'isica Te\'orica y Computaci\'on\\
}

\date{Accepted XXX. Received YYY; in original form ZZZ}

\pubyear{2020}

\begin{document}
\label{firstpage}
\pagerange{\pageref{firstpage}--\pageref{lastpage}}
\maketitle

\begin{abstract}
We have developed a new tool to analyse galaxies in the EAGLE simulations as close as possible to observations. We investigated the evolution of their kinematic properties by means of the angular momentum proxy parameter, \lamRe\, for galaxies with \Mstar\,$\ge5\times10^9$\,\Msun\ in the RefL0100N1504 simulation up to redshift two ($z$\,=\,2).  Galaxies in the simulation show a wide variety of kinematic features, similiar to those found in  integral-field spectroscopic studies. At $z$=0 the distribution of galaxies in the \lamReeps\ plane is also in good agreement with results from observations. Scaling relations at $z$\,=\,0 indicate that there is critical mass, M$_{\mathrm{crit}}$\,/\Msun$=10^{10.3}$ , that divides two different regimes when we include the \lamRe\ parameter. The simulation shows that the distribution of galaxies in the \lamReeps\ plane evolves with time until $z$\,=\,2 when galaxies are equally distributed both in \lamRe\ and $\varepsilon$. We studied the evolution of \lamRe\ with time and found that there is no connection between the angular momentum at $z$\,=\,2 and $z$\,=\,0. All systems reach their maximum \lamRe\ at $z$\,=\,1 and then steadily lose angular momentum regardless of their merger history, except for the high star-forming systems that sustain that maximum value over time. The evolution of the \lamRe\ in galaxies that have not experienced any merger in the last 10\,Gyr can be explained by their level of gas accretion.
\end{abstract}




\section{Introduction}

Integral Field Spectroscopy (IFS) has become the standard tool to perform spectroscopic analysis and study the internal structure of galaxies. This technique allows to perform a detailed analysis of the kinematic and stellar populations over a two dimensional field-of-view, which is crucial to understand the history of individual galaxies and substructure, and disentangle different formation scenarios \citep[e.g.][]{emsellem2004velmaps,emsellem2011atlas3d,Pinna2019A&A...625A..95P}.

The SAURON project \citep{Bacon.SAURON.2001MNRAS.326...23B} was one of the first works that showed the potential of this technique by studying 48 early-type galaxies. This survey revealed that the observed distribution of stars is not strongly correlated with the stellar kinematics, and that galaxies can be divided as Fast and Slow rotators depending on their level of rotational support \citep[e.g.][]{emsellem2007sauron}. This classification was based on the \lamRe\ parameter, developed by the same team as a proxy for the projected angular momentum. Fast and Slow rotators were found to be physically different groups of galaxies, the latter being predominantly massive galaxies. These results were confirmed by the ATLAS$^{3D}$ survey \citep{CappellariAtlas3D.2011MNRAS.413..813C} with an extended sample of 260 early-type galaxies \citep{emsellem2011atlas3d, Krajnovic.KinMorph.2013MNRAS.432.1768K}. Since then, a statistically significant number of galaxies have been studied confirming the early results of SAURON and ATLAS$^{3D}$, providing more information about the different nature of Slow and Fast rotators. Some of the most important surveys are: CALIFA \citep{Sanchez.CALIFA.2016A&A...594A..36S} which studied a sample of $\approx$ 600 galaxies across the Hubble sequence, SAMI \citep{Brough.SAMI.2017ApJ...844...59B} which analysed around 3,000 galaxies across a large range of environments, MASSIVE \citep{Veale.MASSIVE.2017MNRAS.464..356V} which focused on a reduced sample of $\approx$100 very massive galaxies in the nearby Universe and MaNGA \citep{Bundy.Manga.2015ApJ...798....7B}, which will complete the observations of $\approx$10,000 galaxies in 2020.

A number of idealized and cosmological zoom-in simulations were developed after the initial findings to understand the formation mechanisms of Slow and Fast rotators. It was shown that mergers remnants can evolve in very different ways and thus, that there is not a unique formation mechanism of Fast and Slow rotators \citep{Jesseit.2009MNRAS.397.1202J,Bois.mergers.2010MNRAS.406.2405B, Naab.2014MNRAS.444.3357N}. Major mergers have a larger impact in the star formation rate (SFR) and kinematic of galaxies than minor mergers and thus are considered as one of the main drivers of Slow rotators. Nevertheless,  they do not always destroy the rotational dominated nature of the primary galaxy and can easily produce a fast rotating remnant. \citep{Sparre.2017MNRAS.470.3946S, Pontzen.2017MNRAS.465..547P}.  The angular momentum is a key parameter to understand not only the kinematic properties of galaxies, but also their evolutionary path.  It is believed that the amount of angular momentum transferred from halo to disk provides constraints to the size of the galactic disk \citep{Mo.Disks.1998MNRAS.295..319M} and to set the basis for the mass-size relation of galaxies \citep{Shen.Size.2003MNRAS.343..978S}. Moreover, the angular momentum provides a possible explanation for some aspects of the observed morphology-density relation, by means of spiral galaxies transformation into fast rotating lenticular galaxies through fading of the stellar population \citep[e.g.][]{Cappellari.environment.2011MNRAS.416.1680C}.

Current IFS surveys can only observe galaxies in the local universe and until the the next generation of telescopes such as the Extremely Large Telescope (ELT) and the James Webb Space Telescope (JWST) are fully functional it is not possible to develop an angular momentum evolution theory entirely based in observations. On the other hand, there are surveys that use IFUs to study ionized gas kinematics at high redshifts by measuring hot gas emission lines with timescales differing in various orders of magnitude \citep{Foster.SINS.2009ApJ...706.1364F,wisnioski2015kmos3d}, but the connection with the stellar kinematics is not straightforward as they are affected by different physical processes. Fortunately, it is possible to study the stellar kinematics of galaxies at different redshifts via state-of-the-art cosmological hydrodynamical simulations, which are able to reproduce a broad variety of environments and follow the evolution of thousands of galaxies. Examples of these simulations include EAGLE\citep{schaye2015eagle}, IllustrisTNG \citep{Pillepich.ILLUSTRISTNG.2018MNRAS.473.4077P}, HORIZON-AGN \citep{Dubois.HORIZONAGN.2014MNRAS.444.1453D} and Magneticum Pathfinder\footnote{http://www.magneticum.org}. These simulations have proven, with different levels of success, to reproduce key observables in the local Universe and have recently started to explore the kinematics of the galaxies and match the results to those of IFS surveys. \cite{lagos2018connection} and \cite{Choi.Horizon.2018ApJ...856..114C} focused on the decrease of angular momentum due to merger events in the EAGLE and Horizon-AGN simulations respectively. \cite{Schulze2018MNRAS.480.4636S} studied the distribution of different kinematic features in early-type galaxies and their angular momentum evolution using Magneticum. \cite{PillepichTNG502019} analysed the evolution of the intrinsic velocity dispersion in disk galaxies with IllustrisTNG. Each of these teams has developed its own methodology to characterize the morphology and kinematics of galaxies and thus is not always straightforward to compare their predictions. In addition, often, they do not consider all the particularities of kinematic extraction in real observations making it sometimes difficult to compare with IFS results.\looseness-2 

The aim of this work is to develop a methodology to analyse simulated galaxies as close as possible to observations, obtaining IFU-like two-dimensional maps of the kinematics and stellar populations of galaxies. We will use the EAGLE simulations as our reference to obtain predictions that can be tested observationally, given its ability to reproduce properties such as the sizes \citep{furlong2016size}, masses and\citep{Furlong.GSMF.2015MNRAS.450.4486F},  angular momentum \citep{lagos2016angular}, colours \citep{trayford2015colours},  gas content \citep{bahe2015distribution,lagos2015molecular} and color magnitude relation \citep{correa2017relation}. On the other hand, the simulations are also known to deviate from some observational results e.g flatter stellar mass-metallicity relation than the one inferred from observations, a z = 0 transition from passive to active galaxies at too high stellar mass \citep{schaye2015eagle,trayford2015colours} and substantially overproduce both intermediate and young stellar populations \citep{2019NatAs.tmp....1S} \looseness-2

The paper is organized as follows. In Section~\ref{sec:EAGLE} and \ref{sec:Sample} we introduce the key aspects of the EAGLE simulation and describe the sample selection. In Section~\ref{sec:Methodology} we present the methodology used to measure the kinematic properties of simulated galaxies. Section \ref{sec:kin_maps} shows the variety of kinematic features displayed by EAGLE galaxies. In Section~\ref{sec:lambda_R_z0}, we study the distribution of galaxies in the \lamReeps\ at redshift zero ($z$\,=\,0) and compare it with results from the CALIFA survey \citep[][]{fb19}. In Section~\ref{sec:redshift} we present the evolution of galaxies in the \lamReeps\ from $z$\,=\,0 to $z$\,=\,2 and connect it with different fundamental parameters. Finally, we summarize and conclude in Section~\ref{sec:conclussion}. 

\section{The EAGLE simulation}
\label{sec:EAGLE}

EAGLE (Evolution and Assembly of Galaxies and their Environments; \cite{schaye2015eagle}; \cite{crain2015eagle}) is a project from the Virgo consortium that consists of a series of cosmological hydrodynamic simulations aimed at understanding the formation and evolution of galaxies in a $\Lambda$CDM Universe\footnote{EAGLE simulations adopt the cosmological parameters extracted from \cite{collabPlanck} ($\Omega_{m}=0.307$, $\Omega_{\Lambda}=0.693$, $\Omega_{b} = 0.04825$, $h =0.6777$,  $\sigma_{8}=0.8288$, $n_{s}=0.9611$, $Y=0.248$)}. The results of the simulations are publicly available at http://icc.dur.ac.uk/Eagle/database.php. They contain: (1) a catalog that includes properties of haloes and galaxies in the simulation \citep{mcalpine2016eagle}, and (2) the particle data \citep{2017publicreleasedata} for 29 snapshots ranging from $z=20$ to $z$\,=\,0, with time spans that range from 0.1 to 1.3\,Gyr. Simulations were performed using an updated version of the N-Body Tree-PM smoothed particle hydrodinamic (SPH) code, GADGET3 described in \cite{springel2005cosmological}. The modifications in this version, collectively referred to as 'Anarchy', include the pressure-entropy formulation of \cite{hopkins2012general}, the time step limiter proposed by \cite{durier2011implementation}, an artificial conduction switch of the form of \cite{price2008modelling} and the artificial viscosity switch of \cite{cullen2010inviscid}.

\begin{figure*}
\centering
\includegraphics[width=\textwidth]{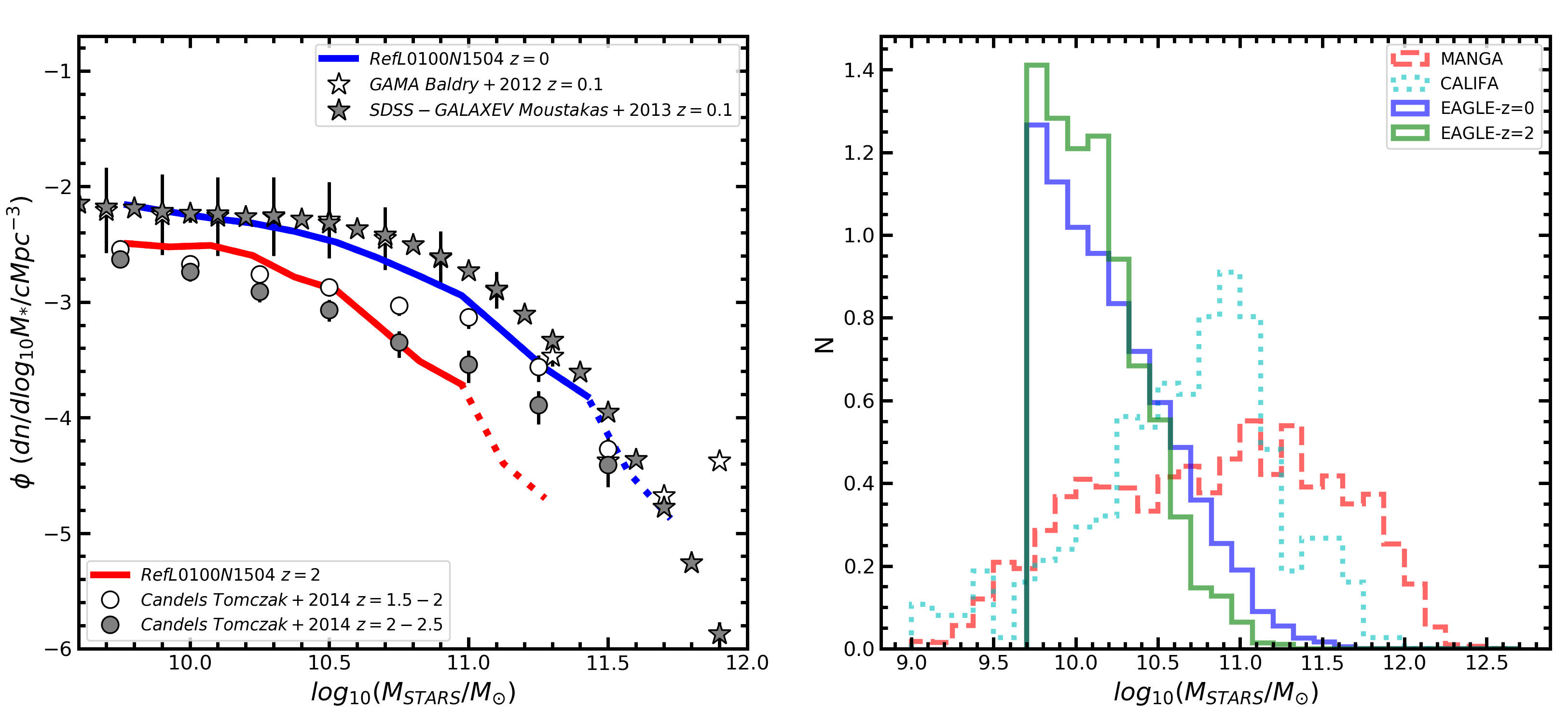}
\caption{{\it Left:} The Galaxy Stellar Mass Function (GSMF) of galaxies in the RefL0100N1504 simulation at $z$\,=\,0 (blue curve) and $z$\,=\,2 (red curve). The curves are dashed for mass bins containing less than 10 galaxies. The GSMF inferred from observations is represented using white \citep{Baldry.GSMF.2012MNRAS.421..621B}  Sand black stars DSS-GALEX \citep{Moustakas.GSMF.2013ApJ...767...50M} at $z$\,=\,0 and white and black circles  \citep{Tomczak.GSMF.2014ApJ...783...85T} at z=2. {\it Right:} Normalized mass distributions of galaxies in the RefL0100N1504 simulation at $z$\,=\,0 (blue) and $z$\,=\,2 (green), in the MaNGA survey(dashed red) and in the CALIFA survey (dotted blue)} 
\label{fig:GSMF_MASS_distribution}
\end{figure*}

Dark matter haloes are detected within the simulation with the Friends-of-Friends (FoF) algorithm \citep{springel2001populating}. Galaxies are then identified as the stellar component of self-bound structures in each dark matter halo detected with the SUBFIND algorithm \citep{dolag2009substructures}. The galaxy that contains the particle with the lowest gravitational potential within a FoF halo is defined as central galaxy, usually the most massive in the halo, and the remaining ones are considered its satellites. 

In order to characterize the evolution of galaxies at different snapshots, merger trees have been created so we can link galaxies with their descendants and progenitors at different times \citep{Qu.MergerTree2017MNRAS.464.1659Q}. To easily identify galaxies and navigate the merger trees, the database provides the \textit{GalaxyID} which is the unique identifier of a galaxy in the simulation, and the \textit{DescendantID} which is the \textit{GalaxyID} of the unique descendant of the galaxy. A galaxy may have multiple progenitors, but only one descendant. An important aspect of EAGLE is the use of sub-grid routines that account for physical processes that act on scales below the resolution limit of the simulations. These sub-grids modules include the radiative cooling and photoheating models of \cite{wiersma2009effect}, the star formation as described by \cite{schaye2008relation}, the stellar evolution and chemical enrichment of \cite{wiersma2009chemical}, stellar feedback of \cite{dalla2012simulating} and a black hole growth and active galactic nuclei similar to that of \cite{rosas2015impact}. Throughout the text, we will focus on the simulations that use the reference model, introduced in \cite{schaye2015eagle}. In this model, the sub-grid parameters that regulate feedback from star formation and black hole accretion were calibrated to guarantee that EAGLE galaxies reproduce observational relations at $z$\,=\,0, such as the galaxy stellar mass function, the galaxy size-stellar mass relation and the black hole mass-stellar mass relation. Additionally, the simulation has been able to reproduce observables such as the color bimodality,  with a blue cloud mostly formed by disky galaxies and a red sequence of predominantly elliptical galaxies \citep{correa2017relation} or the density of complex systems such as ring galaxies \citep{Elegali.RINGS.2018MNRAS.481.2951E}. This proves the success of the numerical model that describe the subgrid physics, since their calibration did not include information about galaxy morphology. \looseness-2

In this work, we will analyse the RefL100N1504 simulation which is characterized by: (1) use of calibrated sub-grid parameters, (2)  simulated cubic volume of 100 co-moving Mpc$^3$ (hereafter cMpc), (3) $2\times1504^3$ initial particles (baryonic and dark matter), (4) initial mass of gas particles, m$_{g}=1.81\,\times\,10^{6}$\,\Msun\ and dark matter particles mass, m$_{dm}$\,=\,9.70\,$\times\,10^{6}$\,\Msun, (5) a co-moving gravitational softening length, $\varepsilon\,=\,2.66$ comoving kpc (ckpc hereafter) for $z \geq 2.8$ and 0.7 proper kiloparsecs (pkpc hereafter) for $z$\,<\,2.8. 
\section{Sample selection}
\label{sec:Sample}

In order to analyse the evolution of galaxies with redshift our sample must include a statistically significant number of objects and cover a large range in mass, \Mstar\,$\ge\,5\times10^{9}\,$\Msun. We will base our target selection on the galaxy stellar mass function (hereafter GSMF) for this purpose. Following \citet{Furlong.GSMF.2015MNRAS.450.4486F} we measured the stellar mass content of galaxies, \Mstar, within spheres of 30 pkpc centred at the minimum of potential of the galaxy. In this way, we include almost all the stellar mass in the subhalo for low mass systems, while for massive systems we exclude the diffuse mass that would contribute to the intracluster light (ICL). In the left panel of Fig.~\ref{fig:GSMF_MASS_distribution} we plot the GSMF of the RefL0100N1504 simulation at $z$\,=\,0 (blue) and $z$\,=\,2 (red) for galaxies with \Mstar\,$\ge\,5\times10^9\,$\Msun\ in mass bins of 0.15 dex. The dotted lines indicate the mass bins where there are less than 10 galaxies. We set the lower mass limit after checking that galaxies with masses below this threshold provided results that did not fulfil our quality requirements (see Sec.~\ref{sec:Methodology}). We compare the GSMF of the simulation with observational results in the left panel of Fig.~\ref{fig:GSMF_MASS_distribution} using GAMA \citep{Baldry.GSMF.2012MNRAS.421..621B} and SDSS-GALEX \citep{Moustakas.GSMF.2013ApJ...767...50M} surveys at $z\approx0.1$ and from ZFOURGE/CANDELS surveys \citep{Tomczak.GSMF.2014ApJ...783...85T} at redshift ranges of $z$\,=\,1.5$-$2 and $z$\,=\,2$-$2.5. The observational works extend down to masses around $10^7$\,\Msun, but we only plot the GSMF in the range of masses in which we are interested.

The GSMF is usually fitted by a Schechter function when a wider range of masses is considered, but for our purposes we visually compare the plots and identify the regions where the simulated and observed values are comparable. At $z$\,=\,0 there is a good agreement with the observational data in the entire mass range of interest, with an average difference of 0.2 dex in the range $10^{9.7} \ - \ 10^{11.5}$\,\Msun. It is remarkable that even at the region of masses $\geq\,10^{11.5}$\,\Msun, where the population of massive galaxies in the simulations is low, theere is good agreement with the observational results. At $z$\,=\,2 our results are also in good agreement with the observations except for masses larger than $10^{11}$\,\Msun. Again, this is because the volume probed by the simulation is too small to have large clusters, where the most massive galaxies inhabit. We therefore confirm that our sample rreproduces the GSMF well for masses $M_\star \geq 3x10^{10} M_\odot$ at $z=0$ and $M_\star \geq 10^{11} M_\odot$ at $z=2$.

We present in Fig.~\ref{fig:GSMF_MASS_distribution} (right panel) a different way to compare our sample's mass distribution with other observations. The figure shows the normalized mass distribution of galaxies in the RefL01001504 simulation with \Mstar\,$\ge\,5\times10^{9}$\Msun\ at $z$\,=\,0 and $z$\,=\,2 (filled blue and green line). In adition we include results for galaxies with angular momentum measurements from the MANGA (dashed line) and CALIFA surveys (dotted line). The number of galaxies that satisfy our selection criteria, $N_{GAL}$, is larger at $z$\,=\,0 ($N_{GAL}=5,587$) than at $z$\,=\,2 ($N_{GAL}=2,523$) as galaxies increase their baryonic content with time through merger events. On the other hand, the MANGA and CALIFA samples consist on 2,300 and 300 objects respectively. The range of stellar masses in the MANGA survey is between $10^{9}-10^{12.5}$\,\Msun\ and galaxies where chosen so the shape of the mass distribution is almost flat in all the mass range $10^{9.75}-10^{12}$\,\Msun. CALIFA galaxies cover almost the same range of masses but the distribution presents a clear maximum at $10^{11}$\,\Msun\ and a smaller fraction of massive galaxies. Given the different shapes of the mass distributions and the lack of massive systems in the simulations the comparison with observations must be carefully handled to prevent biases.

\section{Methodology}
\label{sec:Methodology}

EAGLE simulations give us access to all the physical parameters associated to the particles that conform the galaxies. For the proper comparison with observations, one must take into account all the instrumental peculiarities and methodologies associated to the observational parameters of interest. To this end, we have developed a new tool, written in \textsc{Python 3}, aimed to measure properties of simulated galaxies as close as possible to what it is done in observations. In this section we summarise the details of our procedure.


We first project the stellar particles of the  target galaxies into 2D maps with a predefined odd number of pixels per side, \textit{$N_{pixel}$}. This defines the field of view (FoV hereafter) used to analyse each galaxy. Galaxies are projected along one of the main axis of the simulation. The projections perpendicular and parallel to the stellar spin vector define the edge-on and face-on projections, respectively. 

In galaxies with net rotation, the stellar spin is a proxy of the direction of the rotation axis and thus it should be parallel to the vertical axis of the two dimensional map in the edge-on projection. There are, however, both internal and external factors that can produce a misalignment between them. The external factors are galaxy mergers and tidal forces that  cause a misalignment between the axis of rotation and the stellar spin even when the galaxy presents net rotation around one axis. Their contribution can be reduced by excluding the outer regions of the galaxy, which are less bounded and hence are more easily influenced. The inner region of galaxies, on the other hand, is often dominated by dispersion rather than rotation and thus the stellar spin is a poor descriptor of the global rotation of the galaxy. For this reason, to calculate the stellar spin we take into account the stellar particles within a spherical aperture of 15\,pkpc radius around the centre of potential (CoP) of the galaxy. This aperture is large enough to reduce the contribution of the inner parts of the galaxy while excluding the outer regions.

We produce mass and luminosity density maps with the particles and characterize its projected morphology (see details in Sec. \ref{sec:photometry}). Each stellar particle in the simulation has a mass around $10^6 M_\odot$ and a specified age and metallicity. Each particle can therefore be modelled as a single stellar population (SSP) with that age and metallicity using the EMILES population synthesis models \citep{vazdekis2016uv}. These models provide predictions for SSP for a wide range of ages (e.g. 0.03$-$14.00\,Gyr) and metallicities (e.g.  0.0001$-$0.04 in mass fractions). We assume a \cite{chabrier2003galactic} initial mass function (IMF), which provide the mass-to-light ratios (M/L) in the Sloan Digital Sky Survey \citep[SDSS][]{SDSS.Photometry.1996AJ....111.1748F,SDSS1.2002AJ....123.3487S} filters. The M/L of each stellar particle is computed by interpolating the EMILES values. The effect of dust obscuration is not included in the SDSS luminosities.

We then generate luminosity-weighted maps for different physical parameters. Example of these maps are the projected mean stellar velocity, $V$, and velocity dispersion, $\sigma$, for the kinematics. The metallicity and age maps that are used to analyse the stellar populations of the galaxies. The luminosity-weighted results in this work are obtained using SDSS $g-$band. We also calculate the mass-weighted version of these parameters to study if there is any hidden relation that can not be easiy traced in the observations. 

We analyse galaxies in three different ways: in random, edge-on and face-on projections. We use a square fielf-of-view (FoV) of 81\,pkpc$^2$ (81 pixels of 1\,pkpc) centred at the CoP of the galaxy.

\begin{figure}
\centering
\includegraphics[width=0.5\textwidth]{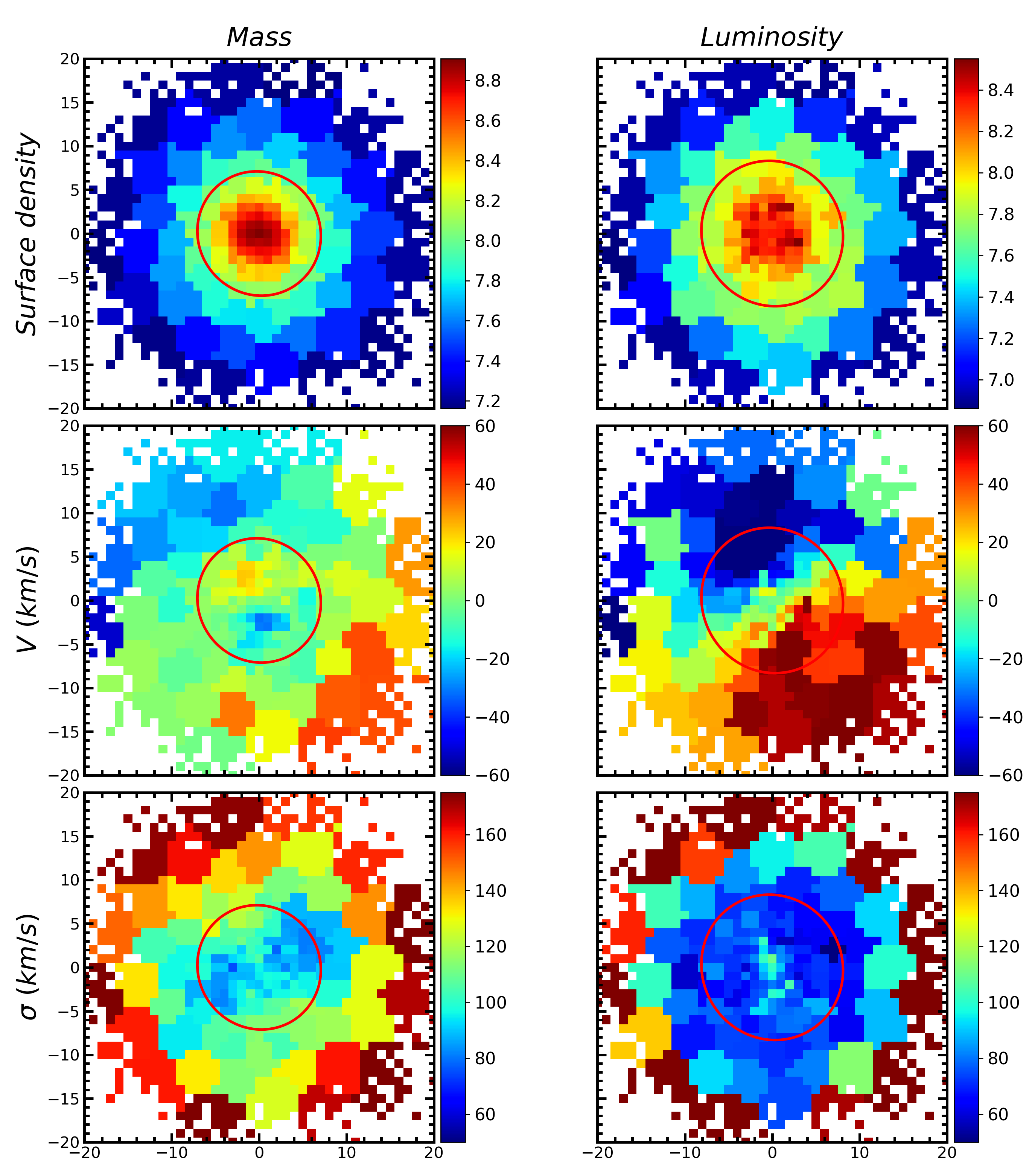}
\caption{Maps of a randomly oriented galaxy in the RefL100N1504 simulation at z=0. Top panel: surface mass density and surface brightness maps. Middle panels: mass and luminosity weighted velocity maps from a Gausian fit to the LOSVD. Bottom panel: mass and luminosity weighted dispersion maps from a Gausian fit to the LOSVD. The  velocities and dispersion maps present the same color range for the mass and luminosity weighted maps. The physical size of the images is shown along the axis and is in pkpc. The red ellipses in the left and right columns are constructed with the morphological information from the maps in the top panel and contain half of the mass and half of the luminosity respectively}
\label{fig:Kin_weight}
\end{figure}

\subsection{Photometric measurements}
\label{sec:photometry}

We use the projected image of the galaxies to construct  surface mass density and surface brightness maps. To that end, we add the mass/luminosity of all the stellar particles in each pixel and divide the total by the pixel area for each pixel in the FoV.  The projected morphology and orientation of the galaxies are characterized by the ellipticity, $\varepsilon$, and position angle of the major axis of the galaxy to the vertical axis, $\theta_{\mathrm{PA}}$, measured counter-clockwise. These parameters are calculated for all projections. We use the \textsc{photutils}\footnote{https://photutils.readthedocs.io} python package to obtain these parameters by diagonalizing the two-dimensional inertia tensor from the surface mass density and surface brightness maps. These measurements can be affected by tidal streams and diffuse haloes of weakly bound stars, and can bias the distribution to smaller ellipticities. To minimize this contribution, we only consider pixels that contain a minimum number of stellar particles, N$_{min}$, thus excluding the outermost regions of the galaxy. Additionally, this step prevents to keep analyzing galaxies with few pixels above the N$_{min}$ requirements. The application of the N$_{min}$ filter is based on the assumption that the distribution of particles in the maps follows a Poisson distribution. Thus, the signal-to-noise ratio (S/N) is equal to the square root of the number of particles in the pixel. We set the N$_{min}$ threshold to 9 particles (i.e. S/N$\sim$3) and we discard galaxies that have less than 10 pixels after the filtering step.

The number of particles per unit area decreases as we move away from the centre of the galaxy and the parameters derived from them consequently loose statistical significance. To ensure that all our measurements are calculated with a minimum number of particles we perform a Voronoi tessellation \citep{cappellari2003adaptive}, which combines pixels into bins that adaptively adjust their size to reach a minimum S/N target (i.e. a minimum number of particles). We set the a minimum S/N per bin of 10 (100 particles). This binning step is applied on the filtered pixels to prevent the formation of excessively large Voronoi bins (VB). In the outermost regions of galaxies the number of stellar particles significantly decrease and with a single VB we would join together stellar particles whose physical properties could be completely unrelated. On the other hand, the binning process may provide an unreliable small number of VBs and thus we impose the following conditions to further analyse a galaxy: (i) it provides a minimum of 10 bins, and (ii) the former condition is satisfied for the random, edge-on and face-on projections.

We calculated the effective radii, R$_{e}$, as the major semi axis of an ellipse with the ellipticity and position angle calculated from the density maps that contain half of the total mass and luminosity. We set a minimum number of 5 VBs inside these ellipses in the three projections to further analyse the parameters derived from the pixels within them. The total luminosity is calculated in the same fashion as the total mass in Sec. \ref{sec:Sample}, taking into account the stellar particles within a spherical aperture of 30\,pkpc of radii centred at the minimum of the potential of the galaxy. We will refer to the radii of the aperture that contains half the mass and luminosity as R$_{e,M}$ and R$_{e,L}$ respectively. The effective radii are computed for each projection.

Most of the kinematic analysis of cosmological simulations use a single effective radius per galaxy regardless of the projection \citep[e.g.][]{lagos2018connection,Schulze2018MNRAS.480.4636S,PillepichTNG502019}. These radii are usually calculated as the radius of the sphere that contains half the mass, or an average of the projected half-mass radius in three directions perpendicular to each other. These parameters provide an estimate of the three dimensional distribution of stars, but are inadequate for comparison with observations, where galaxies can only be studied in the projection along the line-of-sight.  Furthermore, the effective radii measured in observations characterize the aperture that contains half the luminosity of the galaxy not half of its mass. The EAGLE database provides the 3D and the average projected values of R$_{e,M}$ using the stellar particles inside spherical and circular apertures of 30 and 100\,pkpc. The comparison of our values with those in the database (not shown here) indicate that the 3D half mass radii in the database are quite similar to our R$_{e,M}$ values, while our R$_{e,L}$ tend to be somewhat larger (i.e. $\sim$30\%).

In order to make results comparable between galaxies, most of the global parameters presented in this work are calculated using the bins inside elliptical apertures with the same shape and orientation as the surface mass density and surface brightness, and semi-major axis R$_{e,M}$ and R$_{e,L}$. In addition, we use the Voronoi binning tessellation to construct new surface mass density and surface brightness binned maps. In this case we take the mass and the luminosity of the stellar particles inside each bin and divide it by the area of the pixels that conform it.\looseness-2 

\begin{figure*}
\centering
\includegraphics[width=0.99\textwidth]{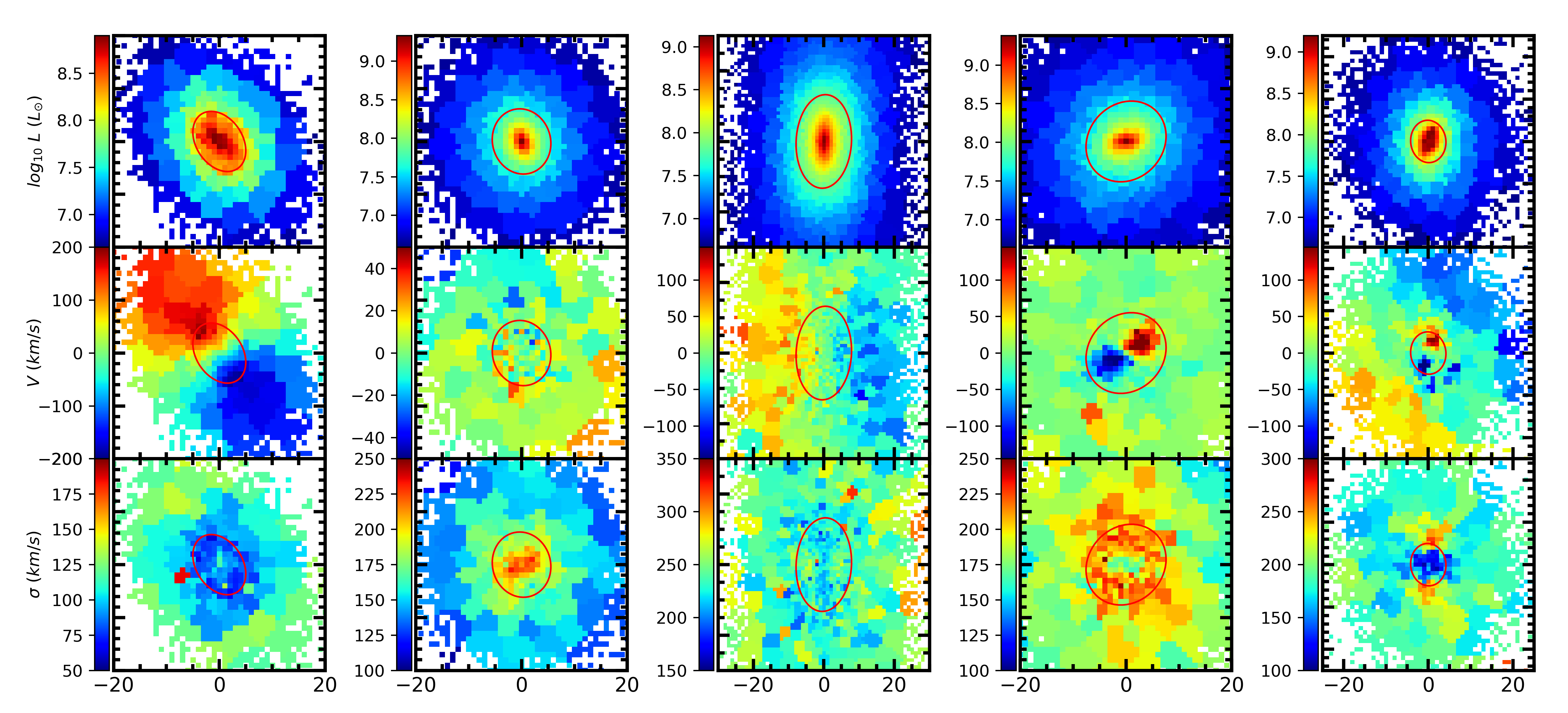}
\caption{Example of surface brightness (top), mean velocity, $V$, (centre) and velocity dispersion, $\sigma$ (bottom) of 5 galaxies that represent the variety of kinematic features in the RefL0100N1504 simulation at $z$\,=\,0. From left to right we have examples of a: regular rotator, non-rotator, prolate rotator, kinematic distinct core and a double $\sigma$ galaxy. The colorbar indicates the scale of each map and the physical size of the images is shown along the axis in pkpc. The red ellipses enclose half the luminosity of the galaxies and are constructed with the morphological information from the maps in the top.}
\label{fig:Kin_features}
\end{figure*}

\subsection{Stellar kinematics}
\subsubsection{Kinematic extraction}

There are two different approaches to calculate the stellar velocity and velocity dispersion from a simulation: \textit{numerically}, where we use directly the velocity of the particles,  and {\it observationally}, where we first obtain the line-of-sight velocity distribution (hereafter LOSVD). There is no consensus in the literature of which one should be applied to simulations when simulations and observations are brought together. 

In the first approach, $V$ and $\sigma$ measurements are directly calculated as the weighted average and standard deviation of the velocities along the line-of-sight at each VB  \citep[e.g.][]{Penoyre2017MNRAS.468.3883P,Schulze2018MNRAS.480.4636S} and \cite{PillepichTNG502019}). In the second approach, we construct the LOSVD in each VB using  the velocities along the line-of-sight, weighting the contribution of each stellar particle by its luminosity. This method is close to what is done with observations using algorithms like pPXF \citep{Cappellari2012pPXF}, where stellar templates are convolved with a LOSVD model to match the galaxy spectra. We adopted the simple approach of a Gaussian distribution to describe the $V$ and $\sigma$ in our galaxies, and defer the computation of higher-order moments of the Gauss-Hermite distributions \citep{van1993new} for future works. In addition, we analyse all the LOSVDs with a fixed size in velocity of 25 km\,s$^{-1}$. The $V$ maps are corrected for the systemic velocity of the galaxy by calculating the mean velocity within a $3\times3$ pixel box around the central pixel.

Some works in the literature use non-weighted versions of $V$ and $\sigma$ to characterize the intrinsic kinematic of galaxies \citep{Schulze2018MNRAS.480.4636S,PillepichTNG502019}. This approach difficults the comparison with observational results since all the information that we get is luminosity weighted. On the other hand, the combined study of mass and luminosity-weighted quantities may provide interesting information about the kinematic of different populations without performing a detailed stellar population analysis. This is because young stars are much brighter and have a larger contribution in the luminosity-weighted parameters, while their mass is in the same range as  their older counterparts. Therefore, the presence of different kinematic features between mass- and luminosity-weighted maps are a quick indicator of young components. 

Figure~\ref{fig:Kin_weight} shows the mass-weighted (left) and luminosity-weighted (right) results for the surface density, $V$ and $\sigma$ maps (top, middle and bottom panels respectively) of a galaxy randomly oriented in the RefL100N1504 simulation at $z$\,=\,0. We see in the $V$ and $\sigma$ maps how the use of different weights completely change the kinematic of the central region. In the mass-weighted $V$ map, we find an example of a counter rotating component where stars in the inner and outer region rotate in opposite directions. There is no sign of such feature in the luminosity-weighted map, indicating that the decoupled component is mostly formed by old stars. This example illustrates that we are likely missing the detection of many of these components simply because of the luminosity-weighting of the observations \citep[see][for further examples]{SAURONVIII.2006MNRAS.373..906M}. In terms of velocity dispersion, there is a region with low $\sigma$ values along the axis of rotation in the mass-weighted map, while in the luminosity-weighted map there is a significant overall drop in the galaxy except for the outermost regions which preserve the high $\sigma$ values. This implies that the young population is distributed all over the galaxy and not only in the inner regions where the counter-rotating component vanishes. The analysis of the role that different populations have on the kinematic properties of galaxies in a global context exceeds the scope of this work and we will only use the luminosity-weighted parameters to compare with observations unless otherwise stated.

\subsubsection{Angular momentum proxy}
\label{sec:lambda_R_defnition}

Galaxies can be classified, based on their dynamical state, into rotational dominated and pressure supported systems. The $V/\sigma$ parameter has been traditionally used with long slit spectrographs taking the central velocity dispersion and the maximum rotational velocity to compute it \citep[e.g.][]{1978MNRAS.183..501B,1983ApJ...266...41D,2005MNRAS.363..937B}. However, with the rise of integral-field units, it has been shown \citep[e.g.][]{emsellem2007sauron} that it failed to differentiate between short-scale rotation (like kinematic decoupled components) and long-scale rotation in early-type galaxies, providing in certain cases very similar values for galaxies with qualitatively and quantitatively different velocity maps. To eliminate this ambiguity, they defined a new parameter called \lamRe\, which is a proxy for the projected stellar angular momentum per unit mass. For the two-dimensional velocity and dispersion maps, \lamRe\ is defined as:
%
\begin{equation}
\label{ec:lambdaR}
\lambda_{R} = \frac{\sum_{j}\omega_{j}R_{j}\mid{V_j}\mid}{\sum_{j}\omega_{j}R_{j}\sqrt{V_{j}^{2}+\sigma_{j}^{2}}},
\end{equation}
%
\noindent where $V_j$ and $\sigma_j$ are the mean velocity and velocity dispersion along the line-of-sight, $\omega_j$ is the surface brightness or mass density of a given bin, and $R_j$ is the galactocentric distance of the $j^{th}$ Voronoi bin, respectively. Purely rotational dominated systems will show \lamRe\ values close to unity while for systems supported by dispersion, with no ordered rotation, this value will be close to zero. To make these measurement comparable between galaxies we only include in the summation of eq. \ref{ec:lambdaR} the VB inside an elliptical aperture with the same $\varepsilon$ and $\theta_{PA}$ as the galaxy and semi-major axis equal to the effective radii.
We adopt here the same criteria defined by  \citet{emsellem2011atlas3d} to separate between Fast and Slow rotating systems (i.e. $\lambda_{R}=0.31\times\sqrt{\varepsilon}$).

\subsection{Galaxy mergers}

We use the merger tree information in the EAGLE database to link galaxies at $z$\,=\,0 with their progenitors at higher redshift. We define the stellar mass ratio between the primary and the secondary galaxy as:\looseness-2
%
\begin{equation}
\label{ec:R_gas_merger}
R_{\mathrm{stars,merger}} = \frac{\mathrm{\Mstar}^{S}}{\mathrm{\Mstar}^{P}},
\end{equation}
%
\noindent where \Mstar\ is the stellar mass of the galaxy and the $P$ and $S$ index represent the primary and secondary galaxy respectively. We classify mergers as: i) major merger for $R_{\mathrm{stars,merger}} \geq$ 0.3, ii) minor merger if $0.1<R_{\mathrm{stars,merger}}<0.3$ and iii) smooth accretion for $R_{\mathrm{stars,merger}}< 0.1$. 

We also divide galaxies into wet (gas rich) and dry (gas poor) mergers using the ratio of the total neutral gas to stellar mass involved in the merger. The amount of neutral  gas (atomic and molecular) is calculated following  \cite{Rahmati.HI.2013MNRAS.430.2427R} which accounts for self-shielding from the galactic ionizing background radiation.  The kind of merger (e.g. wet or dry) is then defined as:
%
\begin{equation}
\label{ec:R_stars_merger}
R_{gas,merger} = \frac{M^{P}_{\mathrm{neutral}}+M^{S}_{\mathrm{neutral}}}{\mathrm{\Mstar}^{P}+\mathrm{\Mstar}^{S}},
\end{equation}
%
\noindent where the new parameter, M$_{\mathrm{neutral}}$, is the neutral gas mass. We then set a limit of $R_{\mathrm{gas,merger}}=0.1$ to separate into wet mergers if they have larger $R_{\mathrm{gas,merger}}$ or dry mergers otherwise. 

We use the previous classifications to get a general stimate of the number and type of mergers, but we notice that they are rather simplistic. They do not properly describe complex scenarios where more than two galaxies merge in the time between two consecutive snapshots. Mergers involving more than two galaxies are analysed as a set of binary mergers, where the principal galaxy is the most massive of the group.

We consider galaxies with \Mstar\,$\ge10^{8}$\,\Msun\ for this measurement to make sure that we are not underestimating the number of minor mergers. Given the large number of galaxies in this extended sample and the high computational cost of gas calculations, we use the mass of star-forming gas available in the database as a proxy for the neutral gas mass. 

\section{Kinematic features}
\label{sec:kin_maps}

We have found that EAGLE galaxies display the wide variety of kinematic features observed in our Universe. After visually inspecting, the velocity and dispersion maps we have visually identified the following kinematic groups:

\begin{itemize}
    \item Regular Rotators (RR): stars rotate around the photometric minor axis with no special features in the velocity map.
    \item Non Rotators (NR): the velocity map shows low-level velocities and no clear rotation axis.
    \item Prolate Rotators (PR): stars rotate around the photometric major axis of the galaxy. 
    \item Kinematic Distinct Component (KDC): the velocity map exhibits two different components in the central and outer parts. Counter-Rotating and misaligned components are included in this group. 
    \item Double $\sigma$ (2$\sigma$): the dispersion map exhibit two off-centre symmetric peaks, with a minimum distance of half the effective radii between them. 
\end{itemize}

\begin{figure*}
\centering
\includegraphics[width=\textwidth]{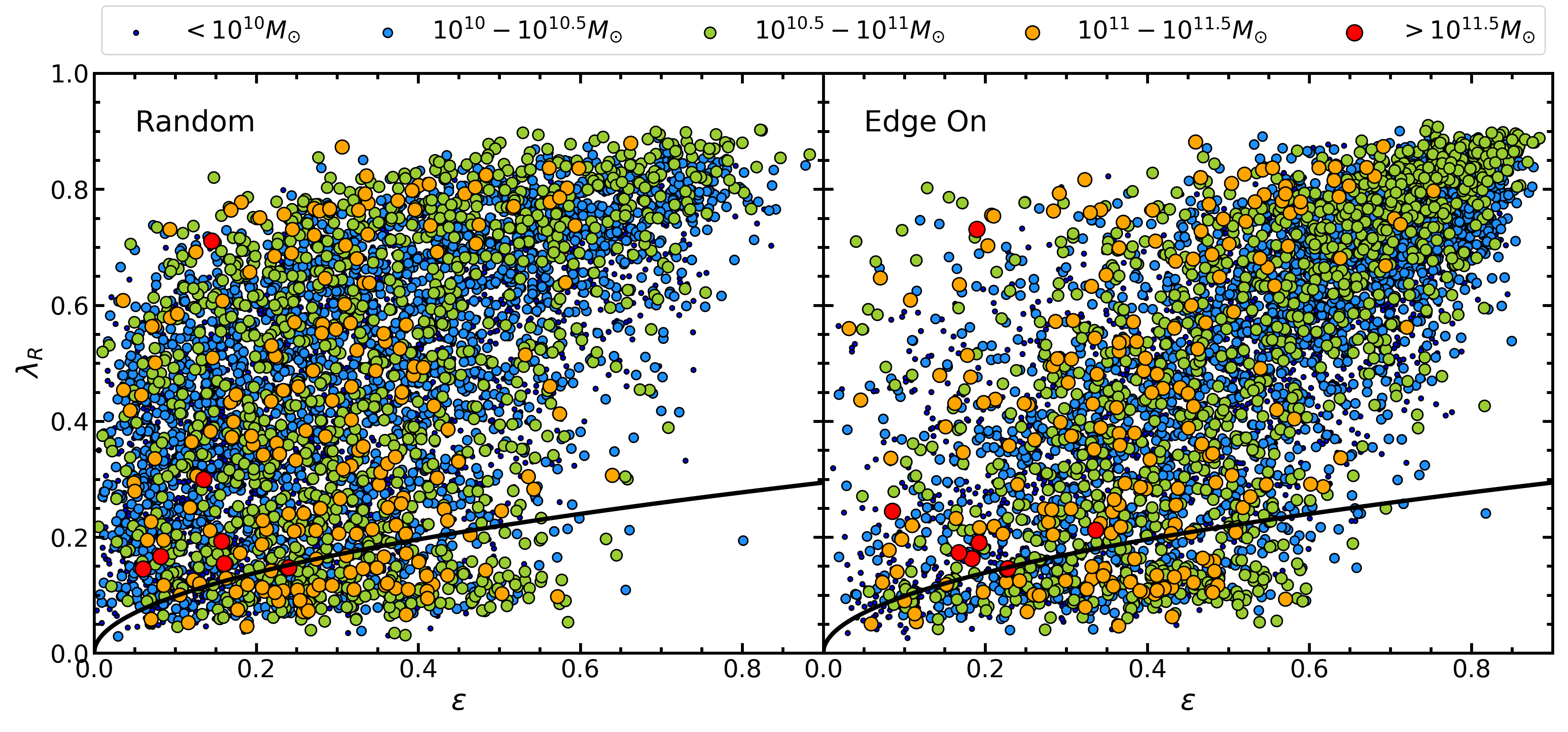}
\caption{\lamReeps\ diagram of EAGLE galaxies with \Mstar\,$\ge5\times10^9$\,\Msun\ in the RefL0100N1504 simulation at z=0. Galaxies are randomly oriented in the left panel and edge-on oriented in the right one. The solid line represent the separation between Fast and Slow rotators from \protect\cite{emsellem2011atlas3d}. The colour and sizes of the symbols scale with the stellar mass of the galaxies as labelled in the top legend.} 
\label{fig:lambda_R_z_0}
\end{figure*}

These are all features already identified in the ATLAS$^{3D}$ survey \citep[e.g.][]{Krajnovic.KineFeatures.2011MNRAS.414.2923K}. In Fig. \ref{fig:Kin_features} we show from top to bottom the surface brightness, mean velocity and velocity dispersion maps for 5 galaxies in the RefL0100N1504 at $z$\,=\,0 with different kinematics features. From left to right we see an example of a regular rotator, a non rotating galaxy, a prolate rotator, a kinematic distinct core and a double sigma peak galaxy. Further, the velocity map of the last galaxy shows a counter rotating core, where the inner region  rotates in the opposite direction of the outer part. Being able to reproduce such a variety of features is a success of the numerical models since no kinematic information is provided to calibrate them. 

Our classification criteria is based on visual inspection and thus a complete kinematic classification of all the galaxies in the simulation exceeds the scope of this analysis. We plan to perform a detailed analysis of the frequency and properties of each kinematic group at different epochs in a future work.

\section{Angular momentum at $z$\,=\,0}
\label{sec:lambda_R_z0}

\subsection{The $\lambda_R - \varepsilon $ diagram}
\label{sec:lambdaR_eps_diagram}

Figure~\ref{fig:lambda_R_z_0} shows the distribution of EAGLE galaxies at $z$\,=\,0 in the \lamReeps\ diagram in random projection (left panel) and in edge-on projection (right panel). The colour and size of the symbols scale with the stellar mass content of the galaxies, so that most massive objects have larger and redder symbols. The solid line indicates the division between Fast and Slow rotators of \cite{emsellem2011atlas3d}. After applying the requisites detailed in \ref{sec:photometry}, the total number of galaxies in these plots is 5,565, a slightly reduced version from the initial sample of 5,587 as some galaxies are excluded from the analysis (see Sec. \ref{sec:photometry}).  In the left panel we observe that galaxies cover a wide region of the parameter space with \lamRe\ ranging from 0.02 to 0.90 and $\varepsilon$ going from 0.03 to 0.88.\footnote{The mass weighted version of Figure~\ref{fig:lambda_R_z_0}  shows a different picture  with almost no galaxies in the simulation with \lamRe\  and $\varepsilon$ larger than 0.8. The latter  implies that our results would deviate more from observations and we could not properly perform the analysis in the following sections.} These values are larger than those in the work of \cite{lagos2018connection}, also based on EAGLE, where galaxies have on average a maximum \lamRe\ equal to 0.8 and $\varepsilon$ equal to 0.65. This is not surprising as  $\varepsilon$ and \lamRe\ were calculated using apertures that contain half the mass that are smaller and thus less sensitive to the global morphology and kinematic of the galaxy. In the right panel, we observe how galaxies move on average to the upper right region of the diagram as they exhibit maximum values of $\varepsilon$ and \lamRe\ in edge-on orientation. We also notice that the number of galaxies showing $\varepsilon<0.3$ and $\lambda_R>0.5$ decrease considerably.

The vast majority of galaxies in EAGLE appear to be Fast rotators, with only 10\% being classified as Slow rotators. The fraction of Slow rotators, $f_{\mathrm{SR}}$, and their dependence with mass and environment has been subject of numerous works in the last years. It has been found that there is a strong connection between the $f_{\mathrm{SR}}$ and the stellar mass \citep{Greene.2017ApJ...851L..33G, Veale.MASSIVE.2017MNRAS.464..356V, Graham.AngularMomentum.2018MNRAS.477.4711G}. This trend was also observed by \citet{lagos2018connection} using the EAGLE and HYDRANGEA simulations. On the other hand, there is also a known relation between Slow rotators fraction and environment, with incresing values towards denser environments \citep{Cappellari.environment.2011MNRAS.416.1680C, DEugenio.2013MNRAS.429.1258D}. Both results are complementary since massive systems are typically found in high-density regions. In fact, \citet{Brough.SR.2017ApJ...844...59B} showed that the majority of Slow rotators are massive and are found in the most overdense regions of clusters. We have calculated the fraction of Slow rotators as a function of the stellar mass and environment using the density indicators defined by \cite{Cappellari.environment.2011MNRAS.416.1680C}. Our results, (not shown here) confirm the increasing trend of the $f_{\mathrm{SR}}$ towards denser environment and higher masses. \citet{lagos2018connection} already addressed this question and confirmed the quick rise of $F_{SR}$ above \Mstar\,$\approx10^{11}$\,\Msun\ using both EAGLE and HYDRANGEA simulations.
\begin{figure}
\centering
\includegraphics[width=0.5\textwidth]{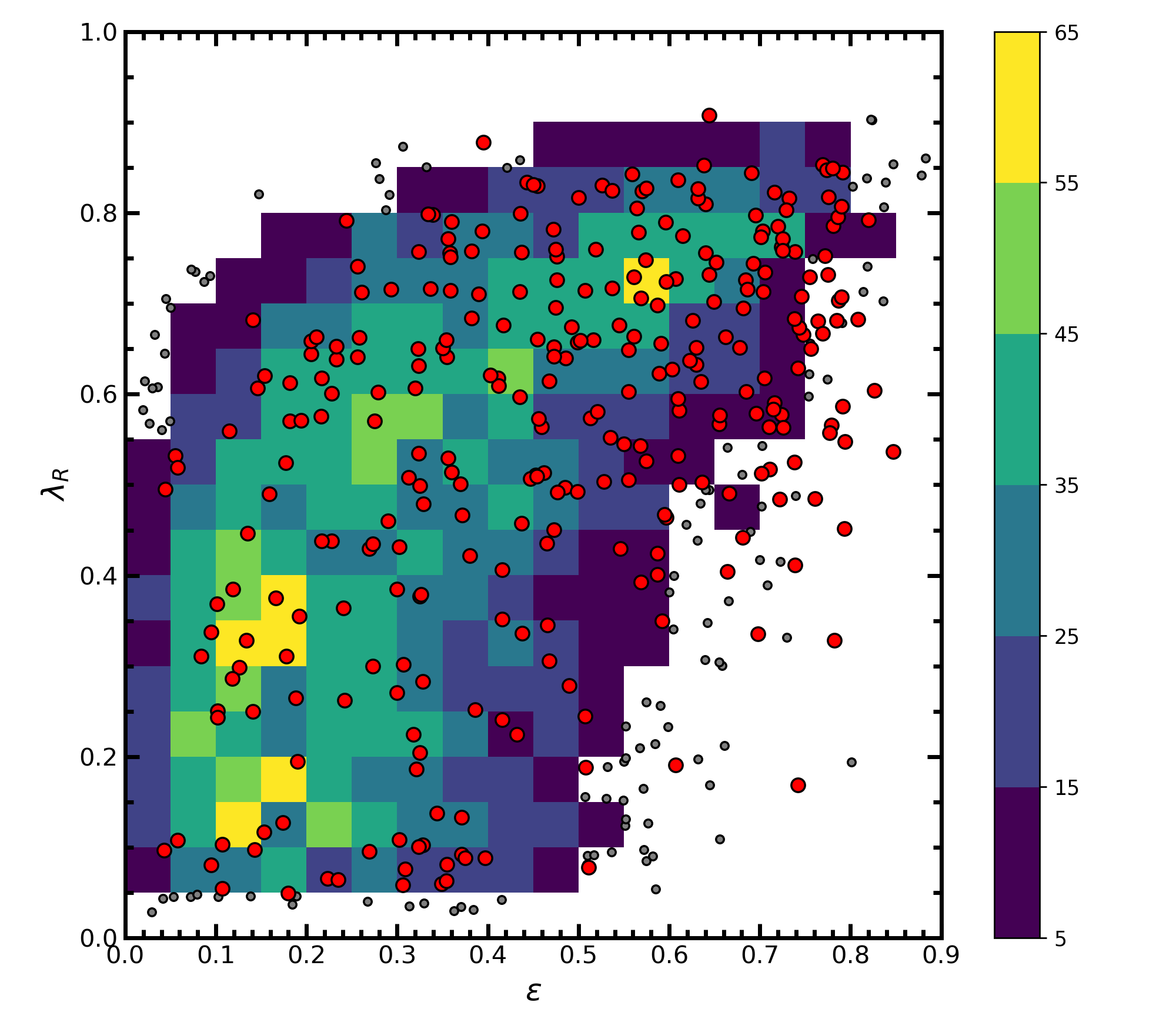}
\caption{The $\lambda_R - \varepsilon$ diagram of galaxies in the CALIFA sample and in the RefL0100N1504 simulation at z=0. The CALIFA sample is plotted using red circles while the colored cells and gray circles represent the simulated galaxies. The color-scale indicates the number of galaxies in each cell and gray points show galaxies in cells containing less than 5 points.}
\label{fig:lambda_R_Califa}
\end{figure}

We now compare our results with the angular momentum measurements of galaxies in the CALIFA survey. This sample consists of 300 galaxies with diverse kinematic properties, from dispersion supported ellipticals, to fast rotating spirals \citep{JESUS.CALIFA.SAMPLE2017A&A...597A..48F}. In addition, the sample covers a wide range of morphologies across the Hubble sequence, but lacks low-mass, low-luminosity early-type galaxies and very massive and luminous late-type systems. Given the large number of galaxies in our sample, we split the \lamReeps\ diagram in 0.05 $\times$ 0.05 cells and color-code them by the number of galaxies they contain. In Fig.~\ref{fig:lambda_R_Califa} we plot CALIFA galaxies in the the \lamReeps\ diagram \citep{fb19} as red circles, while we use the coloured two dimensional grid to represent EAGLE galaxies. To get an estimation of the densest regions of the diagram we only colour the cells that contain a minimum of 5 galaxies in them. Galaxies that fall in the cells with less than 5 galaxies are plotted using gray circles. There is a general good ageement in the range of \lamRe\ and $\varepsilon$ values displayed by both the CALIFA and EAGLE samples. Nevertheless, the specific number of galaxies in each bin is more susceptible to the particular selection functions of each sample.  We note, however, that the lack of extreme flattened galaxies in the EAGLE sample (compared to CALIFA) may be due to the specific isterstellar medium (ISM) subgrid physics used in the simulations. The ISM model used in EAGLE imposes a temperature floor that prevents metal rich gas from cooling below 8000\,K. This temperature threshold sets a minimum disc height of $\approx$\,1\,kpc while Milky Way-like spiral galaxies with large $\varepsilon$ exhibit scale heights of $\approx$\,0.4\,kpc \citep{Kregel.DiskHeigh.2002MNRAS.334..646K}.\looseness-2

\subsection{Relations with fundamental physical properties}

We explore the relation between some fundamental galaxy properties (e.g. mass, size, SFRs and luminosity weighted age) and their connection with the \lamRe\ parameter at $z$\,=\,0.  To calculate the SFR of a galaxy we sum the SFRs of the gas particles that are gravitationally bounded to its sub-halo. Individual SFRs are calculated assuming that the SFR depends on pressure rather than density \citep{schaye2008relation}. This allows to rewrite the observed star formation law of Kennicutt–Schmidt \citep{1998ApJ...498..541K} as a pressure law where the free parameters can be obtained from observations.The great majority of star formation takes place at the central regions of galaxies and the application of the 30\,pkpc spherical aperture has little impact in the final outcome. Ages are calculated as luminosity-weighted averages obtained from the projected age maps in the same apertures used to calculate \lamRe . Galaxies were randomly projected to obtain the values of effective radii, age and \lamRe\.. We use as observational reference the results of the  ATLAS$^{3D}$, CALIFA,  MASSIVE and SAMI surveys shown in \cite{vanSande2019SAMI} . 

\begin{figure}
\centering
\includegraphics[width=0.475\textwidth]{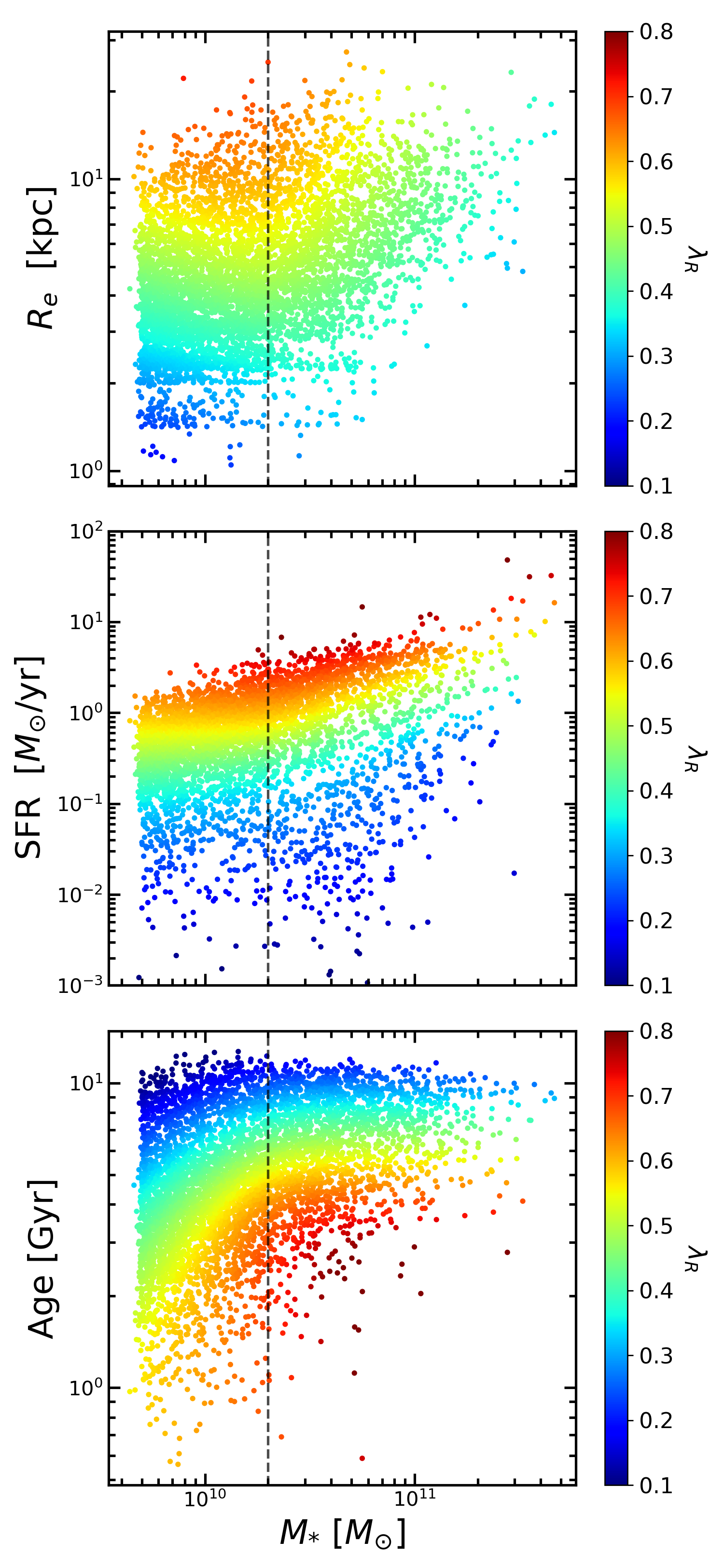}
\caption{Distribution of effective radii (top), star formation rate (midle) and age (bottom) as a function of stellar mass for galaxies in the RefL0100N1504 simulation at $z$\,=\,0. The colorscale indicates the LOESS average \lamRe\ (see text for details).}
\label{fig:plot_LOESS}
\end{figure}

Figure~\ref{fig:plot_LOESS} shows the size (top panel), SFR (middle panel) and age (bottom panel) of EAGLE galaxies as  function of the  stellar mass. The points of these distributions have been colored by their \lamRe\ value. Instead of using the individual values of \lamRe\, we obtained new averaged values to highlight the global trends with the angular momentum. We calculated this new \lamRe\ values using the \textit{CAP\_LOESS\_2D} \footnote{https://www-astro.physics.ox.ac.uk/~mxc/software/} routine of \cite{LOESS.2013MNRAS.432.1862C}, which implements the multivariate, locally weighted regression (LOESS) algorithm of \cite{LOESS.ORIGINAL.doi:10.1080/01621459.1988.10478639}. 

The distribution of galaxies in the size-mass plane is one of the most studied scaling relations, in which most massive systems also display the largest sizes  \citep[e.g,][and references therein]{mass.size.2014ApJ...788...28V}. In the first panel, we show that EAGLE galaxies are on average in good agreement with the observations but are, in general, too large for their mass \citep[see Fig. 3 of][]{vanSande2019SAMI}. This is expected as subgrid modules are calibrated using half-mass radii, which are typically smalles than half-luminosity radii (see Section~\ref{sec:photometry}).  We find that there is a critical value of mass, M$_{\mathrm{crit}}$\,/\Msun$=10^{10.3}$ \ that separates two distinct regimes. Galaxies with masses below M$_{crit}$ have fairly constant sizes at fixed value of angular momentum, while this relation is much steeper (i.e. increasing size with larger \lamRe) for more massive systems.

In the middle panel, we can identify the star forming main sequence for galaxies with  SFR\,$\ge$\,10${-1}$\,\Msun\,yr$^{-1}$, where more massive galaxies tend to exhibit larger values of SFR \citep[e.g.][]{2010ApJ...721..193P}. The good agreement between EAGLE galaxies and observational results of this relation has already been addressed by \citet{Furlong.GSMF.2015MNRAS.450.4486F}. There is a strong correlation between the SFR and the angular momentum, since stars mostly form in in rotationally-supported disks. There is a weak trend with the stellar mass for galaxies with masses larger than M$_{crit}$, where at a fixed value of \lamRe\ more massive galaxies have larger SFRs.

In the bottom panel, we observe that our galaxies span a very similar range of ages as found in observational surveys \citep{vanSande2019SAMI,2015MNRAS.448.3484M,2015A&A...581A.103G} . We find that galaxies with masses below M$_{crit}$ span the largest range of ages from 0.4 to 12.3\,Gyr. There are very few galaxies with ages $\le$\,3\,Gyr at stellar masses larger than $10^{10.5}$\,\Msun. Our measurements strongly deviate from the EAGLE ages shown by \cite{vanSande2019SAMI}, as they showed a considerably older distribution of ages with a minimum age of 2.5\,Gyr and median age of 8.96\,Gyr. We noticed, however, that if we use mass-weighted ages our results are very similar. The distribution of angular momentum in this diagram shows two clearly distinct trends below and above M$_{crit}$. For galaxies with masses lower than $10^{10.3}$\,\Msun\, more massive objects tend to be older at fixed value of \lamRe . For masses larger than  M$_{crit}$ the relation flattens, i.e. at fixed value of age \lamRe\ remains fairly constant.

\begin{figure*}
\centering
\includegraphics[width=\textwidth]{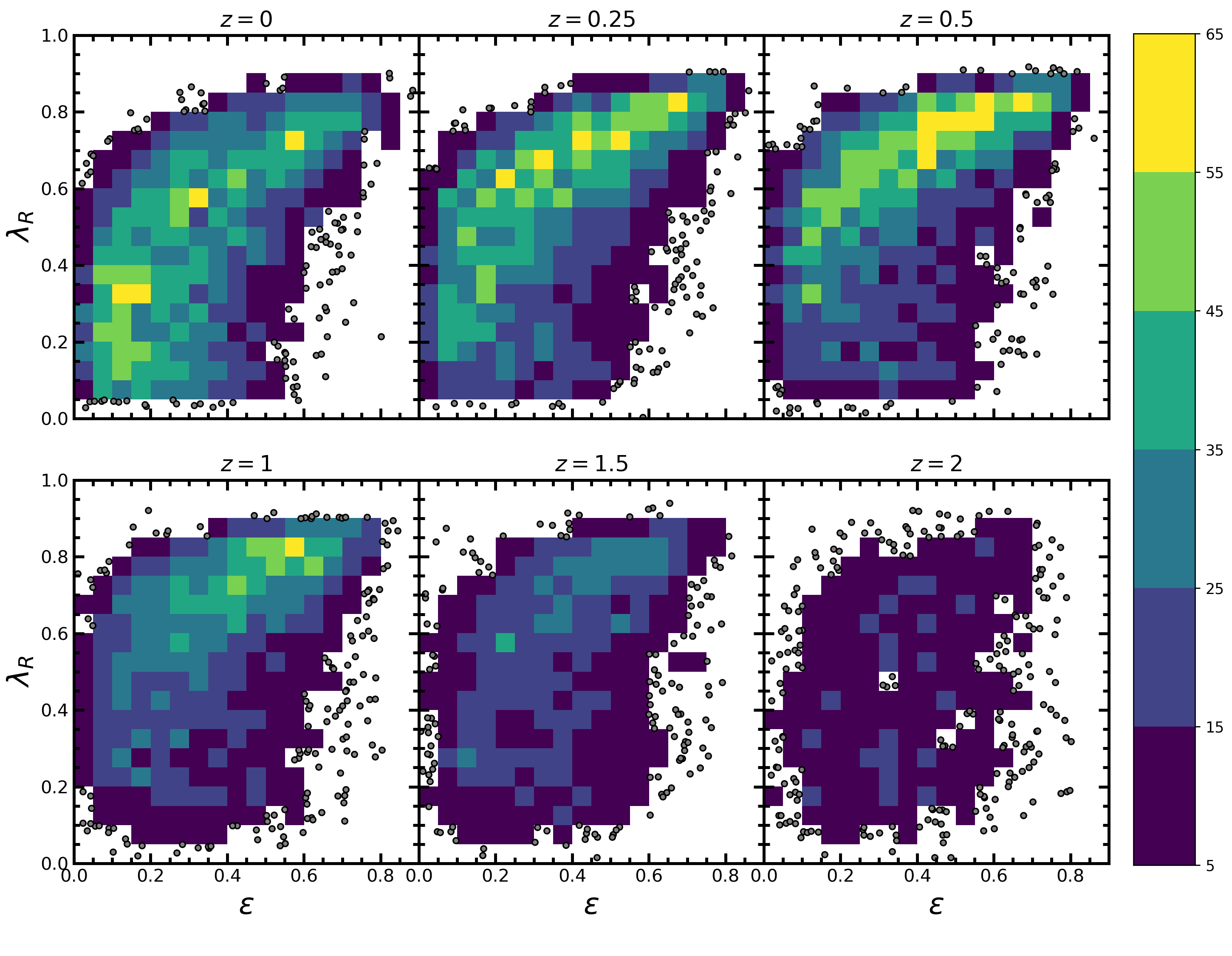}
\caption{$\lambda_R- \varepsilon$ diagram of galaxies with \Mstar\,$\ge5\times10^9$\,\Msun\ in the RefL0100N1504 at $z$\,=\,0, 0.25, 0.5, 1, 1.5, 2 from left to right and top to bottom. The color-scale indicates the number of galaxies in each cell and gray points show galaxies in cells containing less than 5 objects.} 
\label{fig:lambdaR_eps_redshift}
\end{figure*}

\begin{table}
	\centering
	\caption{List of redshift (column 1), number of galaxies in the initial sample (column 2) and number of galaxies successfully analysed (column 3) in the RefL0100N1504 simulation for the snapshots analysed in Sec.~\ref{sec:redshift}}
	\label{tab:sample}
	\begin{tabular}{lccr} 
		\hline
		$z$ & $N_ {\rm sample}$& $N_{\rm analysed}$\\
		\hline
		0.00 & 5,587 & 5,565 (99\%)\\
		0.25 & 5,494 & 5,442 (99\%)\\
		0.50 & 5,345 & 5,193 (97\%)\\
		1.00 & 4,665 & 4,258 (91\%)\\
		1.50 & 3,567 & 3,002 (84\%)\\
		2.00 & 2,523 & 1,874 (74\%)\\
		\hline
	\end{tabular}
\end{table}

\section{Redshift evolution}
\label{sec:redshift}

In this section we focus on the evolution of the dynamical state of galaxies with time. To that end, we analyse the galaxies in the simulation at redshifts $z$\,=\,0, 0.25, 0.5, 1, 1.5 and 2. Table~\ref{tab:sample} indicates the number of galaxies that initially formed our sample at each snapshot and the number of galaxies that are successfully analysed. Due to our mass selection criteria the number of galaxies in our sample decreases with increasing redshift, and we cannot study the progenitors of all the galaxies in our sample at $z$\,=\,0.\looseness-2

The first question that we want to address is how the distribution of randomly oriented galaxies changes with time in the $\lambda_R-\varepsilon$ diagram. Figure~\ref{fig:lambdaR_eps_redshift} shows the $\lambda_R-\varepsilon$ diagram at different redshifts. We proceed in the same way as in Sec.~\ref{sec:lambdaR_eps_diagram} and split the diagram in cells of 0.05$\times$0.05, and color code them by the number of galaxies in each cell. Cells containing less than 5 objects are not coloured and galaxies are indicated using gray circles. We observe that the distribution at $z$\,=\,0.25 is very similar to that of the $z$\,=\,0 diagram. The main difference between these plots is that at $z$\,=\,0 the densest region of the distribution is in the $\lambda_R<0.4$ and $\varepsilon<0.2$ region; whereas at $z$\,=\,0.25 the densest part of the diagram is the region with $\lambda_R>0.5$ and $\varepsilon>0.2$. This trend towards larger values of angular momentum and ellipticity continues at $z$\,=\,0.5 and the most densely populated part of the diagram is now in the region $\lambda_R>0.6$ and $\varepsilon>0.5$. The fourth panel, we notice that at $z$\,=\,1 the number of galaxies in each bin has globally decreased, but the densest region is the same than at $z$\,=\,0.5. At $z$\,=\,1.5 the galaxies are more homogeneously distributed, but the upper part of the diagram still shows a denser region in $\lambda_R>0.6$ and $\varepsilon>0.5$ part. At $z$\,=\,2 galaxies are homogeneously distributed over the entire diagram. It is interesting that we can find galaxies in the $\varepsilon>0.7$ region with a wide range of \lamRe\ values at all redhsifts. We studied the effect of potential biasses due to our selection criteria and they appear to have no effect in these results. \looseness-2

These plots presents and scenario where galaxies are initially homogeneously distributed in the \lamReeps\ diagram at $z$\,=\,2, increase their angular momentum and ellipticity populating preferentially the upper right part of the diagram at $z$\,=\,1 and up to $z=0.25$. At $z$\,=\,0 galaxies appear to experience, on average, a loss of angular momentum and ellipticity, as the bottom left part of the diagram becomes the most densely populated. 

\begin{figure*}
\centering
\includegraphics[width=\textwidth]{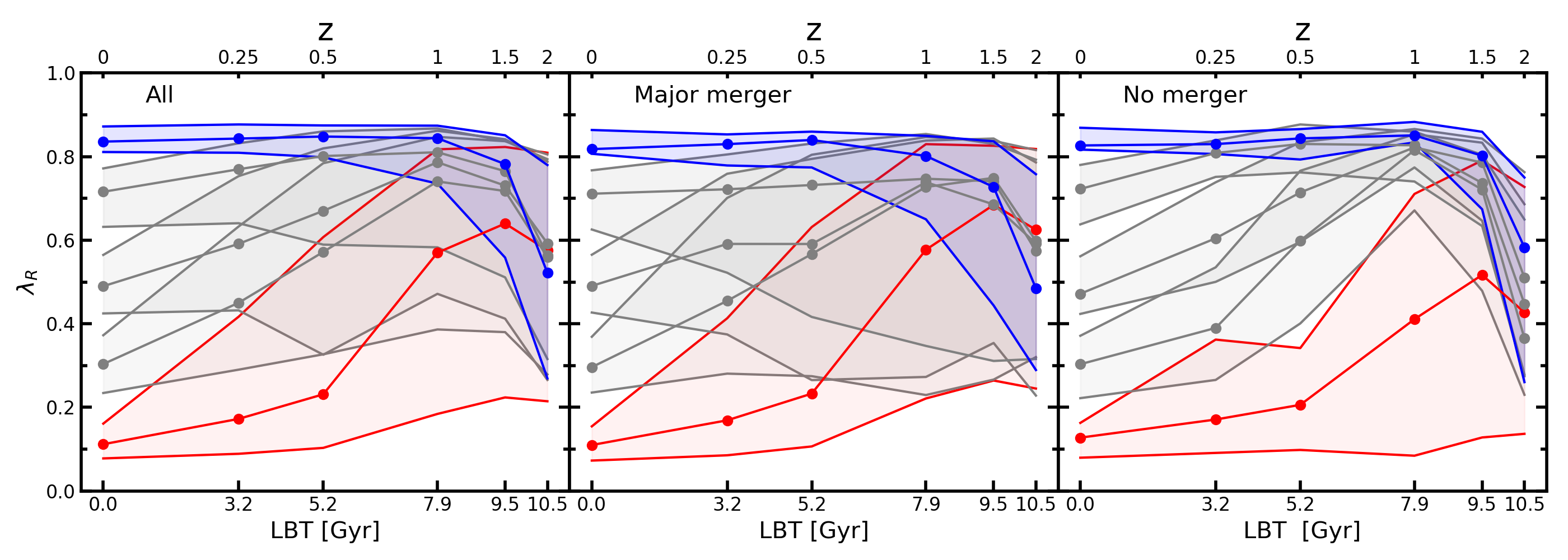}
\caption{Evolution of \lamRe\, as function of look-back time (LBT) and redshift for the five groups of galaxies in \ref{tab:Groups_evolution_mergers}. Circles represent the average value for each group and the lower and upper lines, the 16 and 84 percentiles respectively. Blue and red colours are used for the groups with present-day $z$\,=\,0 values of  $\lambda_R > 0.8$ and $\lambda_R < 0.2$. From left to right the panels show the evolution of all the galaxies, galaxies that experienced a major merger and galaxies that had no mergers.} 
\label{fig:lambdaR_LBT}
\end{figure*}

Motivated by these results, we have analised the evolution of \lamRe\ with redshift for different families of present-day galaxies. In particular, we have explored to what extent this evolution has been dominated by external or secular processes, and whether that determines the angular momentum of galaxies at $z$\,=\,0. In order to fully exploit the cosmological side of the simulation, we focus on galaxies whose progenitors already satisfy our requirements of mass and data quality at $z$\,=\,2, leading to a reduced sample formed by 1,152 galaxies. We also chose galaxies in edge-on orientations to simplify the analysis and avoid biases among galaxies. We divided our sample into five groups of increasing \lamRe\ in steps of $\Delta\lambda_{R}=0.2$ at $z$\,=\,0. We further divide each group according to their merger story.  We show in Table~\ref{tab:Groups_evolution_mergers} the total number of galaxies per group and the number of those that have experienced any major merger, only minor mergers, or had no mergers from $z$\,=\,2. The number of galaxies that only experienced smooth accretion is not indicated  to  facilitate understanding Table~\ref{tab:Groups_evolution_mergers}  but can be calculated straightforwardly with the information therein.

\subsection{The impact of major mergers}

We observe a clear trend where the number of galaxies that have undergone a major merger increases towards lower values of \lamRe\. This is not surprise as major mergers are expected to heavily affect the kinematics of galaxies and produce slow rotators \citep{Bois.mergers.2010MNRAS.406.2405B,Naab.2014MNRAS.444.3357N}. However, the fraction of present-day galaxies with high \lamRe\ that has suffered a major merger is not negligible, and shows that they do not always destroy the rotationally dominated nature of the galaxy. Zoom in simulations have already shown that major mergers can quench the star formation of a galaxy and preserve their disc dominated structure \citep{Pontzen.2017MNRAS.465..547P,Sparre.2017MNRAS.470.3946S} and that even the remnant of a dry major merger can evolve into a fast rotator by disk regrowth if the surrounding gas halo cools down \citep{Moster.Sim.2012MNRAS.423.2045M}. In fact, it was already shown by \cite{lagos2018connection} that very gas rich, wet major mergers with $R_{\rm gas,merger}>0.8$ can spin-up galaxies.

Figure~\ref{fig:lambdaR_LBT} shows the evolution of \lamRe\ as a function of the look back time (LBT) and redshift for all the galaxies (left), galaxies that have experienced any major merger (middle) and galaxies that had no merger (right), divided into the five groups described before. Symbols represent the average value and the contour the 16 and 84 percentiles. Since we want to concentrate our study on the \lamRe\ extreme groups (e.g. with $\lambda_R>0.8$ and <0.2), we use blue and red colours respectively to highlight them, while using gray colour for the rest. On the left panel, it appears that there is no link between the angular momentum at $z$\,=\,2 and $z$\,=\,0. All the five groups present the same average value of $\lambda_{R}\approx0.6$ and spam a large range of values at $z$\,=\,2. Then, all the groups increase their mean angular momentum reaching their peak value between redshifts 1 and 1.5 to then evolve separately down to $z$\,=\,0. The blue group already shows high mean value of \lamRe\ and small scatter at $z$\,=\,1 and is able to maintain it until $z$\,=\,0. The rest of the groups steadily loose angular momentum from $z$\,=\,1 to $z$\,=\,0 with varying values of scatter. In particular, the red group has already started to loose angular momentum at $z$\,=\,1 and it shows large scatter values at all $z$. The middle panel reveals that galaxies that have experienced major mergers are one of the greatest source of scatter in the left panel, since the dispersion in \lamRe\ values for the gray groups is considerably larger. It is interesting that the evolution of the blue and red groups is very similar in both panels. The right panel shows that galaxies that have not experienced any merger evolve in the same way as those that have suffered them. Therefore there has to be a different dominant mechanism that explains the decrease of \lamRe\ with time.

\begin{table}
	\centering
	\caption{Number of galaxies in different kinematic groups. Column 1 Selection criteria . Column 2 Total number of galaxies. Column 3 Number of galaxies that had any major merger. Column 4 number of galaxies that had only minor mergers. Column 4 Number of galaxies that had no mergers.}
	\label{tab:Groups_evolution_mergers}
	\begin{tabular}{lcccr} 
		\hline
		$\lambda_{R,edge-on}$ & $N_{total}$ & $N_ {major}$& $N_{minor}$& $N_{no merger}$\\
		\hline
		>0.8 & 146 & 41 \ (28.1\%) & 51 (34.9\%) & 12 (8.2\%)\\
		0.6-0.8 & 279 & 118 \ (42.3\%) & 57 (20.4\%) & 28 (10.0\%)\\
		0.4-0.6 & 247 & 98 \ (39.7\%) & 59 (23.9\%) & 26 (10.5\%)\\
		0.2-0.4 & 223 & 96 \ (43.0\%) & 43 (29.3\%) & 28 (12.6\%)\\
		<0.2 & 257 & 170 \ (66.1\%) & 30 (11.7\%) & 19 (7.4\%)\\

		\hline
	\end{tabular}
\end{table}

\begin{figure*}
\centering
\includegraphics[width=\textwidth]{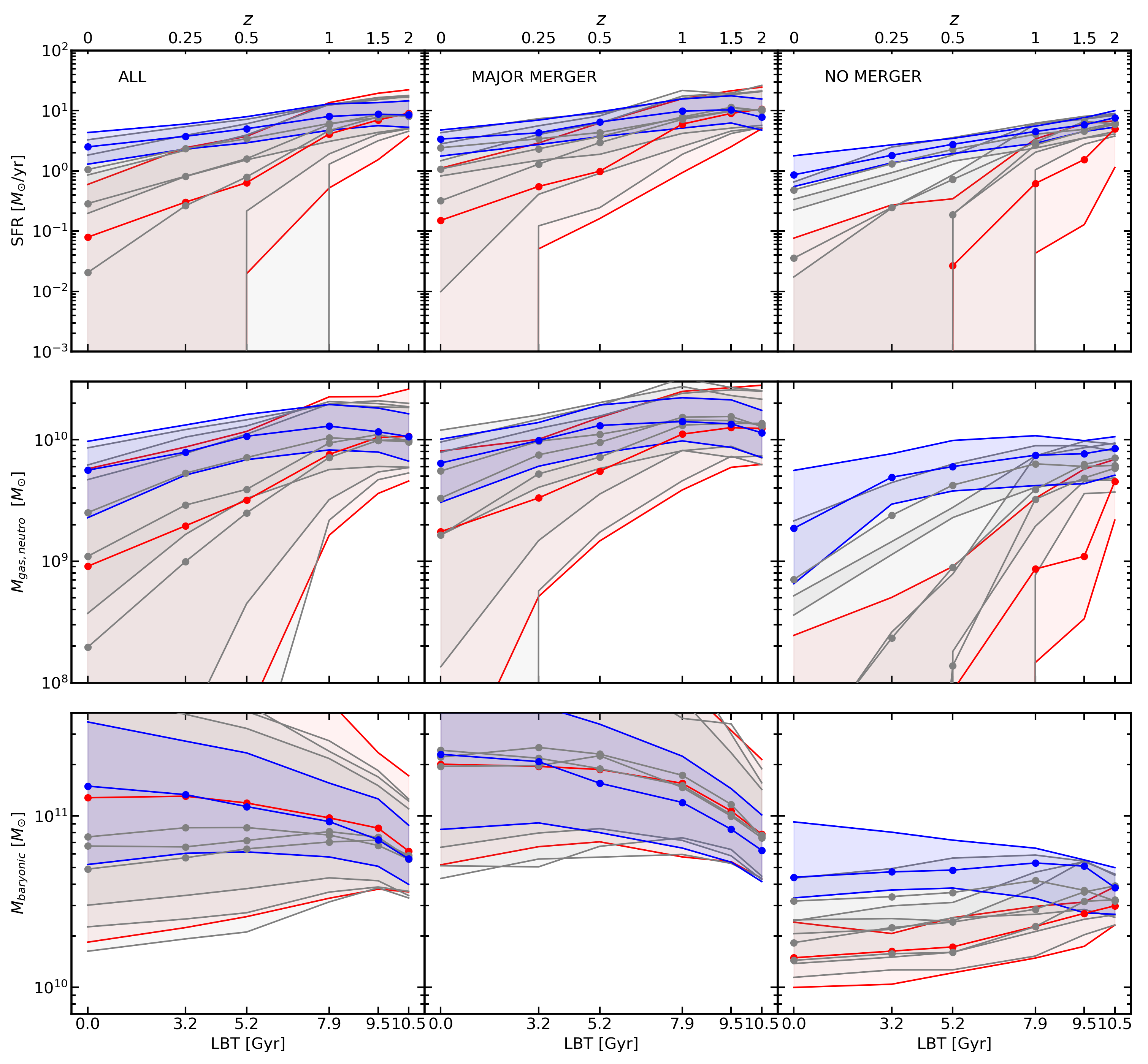}
\caption{From top to bottom, the evolution of the SFR, mass of neutral hydrogen, and baryonic mass as function of the LBT and redshift for the five groups of galaxies in \ref{tab:Groups_evolution_mergers}. Circles represent the average value for each group and the lower and upper lines, the 16 and 84 percentiles respectively. Blue and red colours are used for the groups with $\lambda_R > 0.8$ and $\lambda_R < 0.2$ and gray for the rest. From left to right each row shoes shows the evolution of all the galaxies, galaxies that experienced a major merger and galaxies that had no mergers.}
\label{fig:PRoperties_vs_LBT}
\end{figure*}

\subsection{The role of star formation and neutral gas content}

We studied the \lamRe\ evolution of galaxies that only experienced minor mergers or smooth accretion and found very similar results. Furthermore, we split our sample into centrals and satellites and obtained the same results. On the other hand, when we separated the merger analysis into wet and dry mergers, we found that galaxies that had experienced wet mergers follow the same trends seen in Fig.~\ref{fig:lambdaR_LBT}, while those suffering dry mergers present different evolutionary tracks of \lamRe\ . In order to understand the origin of this phenomenon, we have studied the evolution of different galaxy physical parameters. Figure~\ref{fig:PRoperties_vs_LBT} shows the evolution of the SFR (top row), mass of neutral hydrogen (middle row) and total mass of baryons (bottom row) as a function of time for each group of galaxies. In each row we plot (from left to right) the evolution for all the galaxies, galaxies that experienced any major merger and galaxies that suffered no mergers. The colours and symbols are the same as in Fig.~\ref{fig:lambdaR_LBT}. 

In the top row, it appears that the blue group presents the largest values of SFR with smallest scatter at all times. Galaxies in the red group, on the other hand, always show the lowest values of SFR and largest scatter. This difference is more evident in the right panel, where galaxies that had no mergers are rapidly quenched already at $z$\,=\,1. Interestingly, the five groups of galaxies that had no mergers experience a faster decrease of SFR than galaxies in the other two plots. The evolution of galaxies that had suffered major mergers is very similar to the global sample. 

The middle row offers a similar picture to the one in the top row. Overall there is a general trend for galaxies to lose neutral gas content over time. In addition, there is a clear difference between galaxies that have experienced major mergers from those that have not. The loss of neutral gas content is larger for galaxies that had no mergers, which may explain the larger decrease of SFR of this group in the upper panel. This is not surprising given the known relation between the mass of neutral hydrogen and SFR \citep{2007ApJ...671..333K,2008AJ....136.2782L,Bigiel.2008AJ....136.2846B}. We also find that there is a marked difference among groups for the galaxies that had no mergers. We see that the gas content of the blue group is almost constant from $z$\,=\,2 to $z$\,=\,1 and then decreases down to $z$\,=\,0, while this decrease is much more abrupt for the other families of objects.  

The correlation between reduced SFR or neutral gas and lower spin is explained as a fading effect. The loss of neutral gas quenches the star formation which means that older stellar populations contribute more to the luminosity-weighted spin and therefore decreases.

The bottom row shows that the baryonic mass of galaxies evolve in different ways in each panel. On the left, the mean value and scatter of the baryonic mass for the blue group increases from $z$\,=\,2 to $z$\,=\,0. For the rest of the groups the scatter is so large that it includes the possibility of mass values lower than the initial mass at $z$\,=\,2. The red group presents the largest scatter at all redshifts. The middle panel displays that the five groups increase their baryonic mass from the $z$\,=\,2 up to $z$\,=\,0. The average value and scatter of the five groups is very similar at all redshifts, except for the blue group which shows smaller values of scatter. The right panel shows an interesting trend  where galaxies in the blue and red groups evolve in opposite directions. The former shows a very similar range of mass at all redshift with a small increase in the mean value from $z$\,=\,2 to $z=1.5$. On the contrary, the baryonic mass of the red group decreases from $z$\,=\,2 to $z=0.5$ and then remains constant down to $z$\,=\,0. These trends are so different that at $z$\,=\,0 there is a clear bimodality in the mass distribution of the blue and red groups. The rest of the groups show intermediate behaviours.

The top and middle plots on the right present and scenario where the evolution of \lamRe\ shown in Fig.~\ref{fig:lambdaR_LBT} is tightly correlated with the SFR and neutral gas content. The top and middle panels show that galaxies in the blue group have a source of neutral hydrogen that is constantly providing new gas as they have the largest values of SFR and mass of neutral hydrogen at all redshifts. Since these galaxies do not experience any merger, the gas must have been accreted from the intergalactic medium (IGM). On the contrary, SFR and mass of neutral hydrogen of the red group rapidly drops showing that these galaxies lack the replenishment of new cold gas. These features could explain the different evolution of \lamRe\ seen in Fig.~\ref{fig:lambdaR_LBT}, as it is expected that the angular momentum grows proportionally with time through newly accreted gas \citep{White.1984ApJ...286...38W}. It is remarkable that, even though the blue group always present large values of SFR, their neutral gas content remains almost constant from $z$\,=\,2 to $z=0.25$.  These facts would lead to a baryonic mass that steadily increases for the blue group and that remains constant for the red one, which slightly differs to our plot in the bottom right of Fig.~\ref{fig:PRoperties_vs_LBT}. Although these galaxies do not experience mergers, they do not evolve in isolation and are affected by processes that can remove part of their baryons such as ram pressure and tidal striping.   Therefore, the combination of accretion and stripping processes  experienced by galaxies in the simulation causes these differences.

Regarding galaxies that experienced mergers, the evolution of galaxies that accreted low mass haloes ($R_{\mathrm{stars,merger}}<0.1$) can also be explained in the same scenario as galaxies that had no mergers since their SFR, mass of neutral hydrogen and baryonic mass evolution is very similar. However, the evolution of the baryonic mass presents larger scatter due to the additional source of stars and gas from the accreted galaxies. The latter, makes the analysis of galaxies that have experienced a major or minor merger even more difficult, as mergers do not only act as large source of gas and stellar mass, but they can quench the star formation in short time scales \citep{Lotz.mergers.1.2008MNRAS.391.1137L,Lotz.mergers.2}.  It is worth noting that even though the remnant of a major merger can evolve in many different ways the blue group still shows the largest values of SFR and mass of neutral hydrogen at almost every redshift.

In addition to the three parameters presented here, we investigated the role of other physical parameters such as the dark matter spin \citep{Bullock.2001ApJ...555..240B}, black hole mass, AGN activity or total dark matter mass, among others, but none of them showed a strong connection with the evolution of \lamRe.
 
\section{Conclusion and discussion}
\label{sec:conclussion}

In this work we present the kinematic results of galaxies in the RefL0100N1504 simulation using a novel approach that allow us to study the simulated galaxies as close as possible to observations. We took special care to characterize the projected morphology to truly account for the global distribution of light and not only the central regions. For each galaxy we were able to obtain IFU-like maps of both the kinematic and  stellar populations.

We found that EAGLE simulations are able to generate galaxies with a wide variety of kinematic features such as: regular rotators, non rotators, KDCs, prolate rotators and 2-$\sigma$ galaxies. This a great success of the subgrid physic modules since no kinematic information is used to calibrate them.

We analysed the distribution of galaxies in the $\lambda_R-\varepsilon$ plane and found that there is a good agreement with results from other IFS. However, the simulation lacks the population of massive slow rotators, because of the limited size of the simulated cosmological volume, which cannot produce a statistically significant population of galaxies with \Mstar$\ge10^{11.5}$\,\Msun.

We studied the dependence between the size, SFR and luminosity-weighted average age of the galaxies with their stellar mass and angular momentum. The size-mass and SFR-mass relations had already been studied and our results confirm those in the literature. Our average age-mass results are in good agreement with the stellar ages derived from various IFU surveys, spanning the same range of values.We notice, however, that our results contrast with those of \cite{vanSande2019SAMI} where the authors showed that galaxies in EAGLE were older than in observations. We found that there is critical mass M$_{\mathrm{crit}}$\,/\Msun$=10^{10.3}$ , that separates different trends with the angular momentum.

We also investigated the distribution of galaxies in the $\lambda_R-\varepsilon$ plane at different redshifts. We found that galaxies tend to exhibit larger values of angular momentum and ellipticity towards higher redshifts up to $z$\,=\,1. At $z$\,=\,2 galaxies display similar distributions in the diagram. Galaxies can be found at any region of the diagram with \lamRe\ and $\varepsilon$ <0.9, at any redshift (up to $z$\,=\,2).

Finally, we studied the evolution of \lamRe\ with time of a reduced sample of 1,152 galaxies divided in five different groups of increasing \lamRe\ and found that there is no connection between the angular momentum at $z$\,=\,2 and $z$\,=\,0. We also found that galaxies increase their angular momentum from $z$\,=\,2 to $z$\,=\,1, a moment when galaxies with lower values of \lamRe\ at $z$\,=\,0 begin to steadily loose angular momentum. We analysed the impact of mergers and found that this behaviour is present for all galaxies whether they experienced mergers or evolved undisturbed. We further probed evolution of the SFR, neutral hydrogen and baryonic mass and found that for galaxies that had no mergers the evolution towards high or low angular momentum is mostly caused by gas accretion from IGM. Galaxies which have a constant source of neutral hydrogen from the IGM have the largest amounts of SFR and the continuous generation of young stars is able to keep the high angular momentum, while galaxies that do not accrete gas are rapidly quenched and lose their angular momentum. In the case of mergers the scatter of the distributions is much larger and is not straightforward to obtain the same conclusion. Further analysis are required to understand the effects of gas accretion, merger events and environment in the angular momentum in a fully cosmological context.

\section*{Acknowledgements}
This work has been supported through the RAVET project by the grant AYA2016-77237-C3-1-P from the Spanish Ministry of Science, Innovation and Universities (MCIU) and through the IAC project TRACES which is partially supported through the state budget and the regional budget of the Consejer\'\i a de Econom\'\i a, Industria, Comercio y Conocimiento of the Canary Islands Autonomous Community. We acknowledge the Virgo Consortium for making their simulation data available. The eagle simulations were performed using the DiRAC-2 facility at Durham, managed by the ICC, and the PRACE facility Curie based in France at TGCC, CEA, Bruyeresle-Ch\^{a}tel.




\bibliographystyle{mnras}
\bibliography{EAGLE_kinematics_PaperI_v1} 



\bsp	
\label{lastpage}
\end{document}